\documentclass{PoS}

\usepackage{amsmath}
\usepackage{amssymb}
\newcommand{\tr}{\mathop{\rm tr}\nolimits}
\newcommand{\V}[1]{{\Vec{#1}}}

\title{Disconnected contributions to D-meson semi-leptonic decay form factors}

\ShortTitle{Disconnected contributions to D-meson semi-leptonic decay form factors}

\author{%
G.~S.~Bali$^a$, 
S.~Collins$^a$, 
R.~Horsley$^b$, 
\speaker{I.~Kanamori}$^a$, 
Y.~Nakamura$^c$,
D.~Pleiter$^{a,\,d}$,
P.~P\'erez-Rubio$^a$,
P.~E.~L.~Rakow$^e$, 
A.~Sch\"afer$^a$,
G.~Schierholz$^f$, 
F.~Winter$^b$, 
J.~M.~Zanotti$^g$
(QCDSF Collaboration)\\
\llap{$^a$} Institut f\"ur Theoretische Physik, Universit\"at Regensburg, 
            D-93040 Regensburg, Germany.\\
\llap{$^b$} School of Physics, University of Edinburgh, 
            Edinburgh EH9 3JZ, UK.\\
\llap{$^c$} RIKEN Advanced Institute for Computational Science, 
            Kobe, Hyogo 650-0047, Japan.\\
\llap{$^d$} J\"ulich Supercomputing Centre, Forschungszentrum J\"ulich,
            D-52425 J\"ulich, Germany.\\
\llap{$^e$} Theoretical Physics Division, Department of Mathematical 
            Sciences, University of Liverpool, Liverpool L69 3BX, UK.\\
\llap{$^f$} Deutsches Elektronen-Synchrotron DESY, 22603 Hamburg, Germany.\\
\llap{$^g$} Special Research Centre for the Subatomic Structure of Matter, School
            of Chemistry \& Physics, University of Adelaide, 
	    South Australia 5005, Australia.

E-mail: \email{issaku.kanamori@physik.uni-regensburg.de}}%

\abstract{%
We calculate the disconnected contribution to the form factor for the
semileptonic decay of a $D$-meson into a final state, containing a flavor
singlet eta meson. We use QCDSF $n_f=2+1$ configurations at the flavor
symmetric point $m_u=m_d=m_s$ and the partially quenched approximation for the
relativistic charm quark. Several acceleration and noise reduction
techniques for the stochastic estimation of the disconnected loop are
tested. }

\FullConference{ The XXIX International Symposium on Lattice Field Theory - Lattice 2011\\
July 10-16, 2011\\
Squaw Valley, Lake Tahoe, California}

\begin{document}

\section{Introduction}
Semileptonic decays of $D$-mesons contain rich physics.
Lattice calculations of the form factors for these decays
are important for the search for hints of new physics through 
the determination of CKM matrix elements.
These form factors have been well-studied on the lattice.
Previously, some of us tested a stochastic method to measure 3-point
functions needed to calculate the semileptonic decay 
form factor \cite{Evans:2010tg}.  
The advantage of stochastic methods is that
we have access to a greater range of momenta at fixed cost.
This enables us to  extract the form factor more reliably from 
results for the three point functions at different momentum transfers.

\begin{figure}[bp]
 \vspace*{-0.5em}
 \hfil\includegraphics[width=0.3\linewidth]{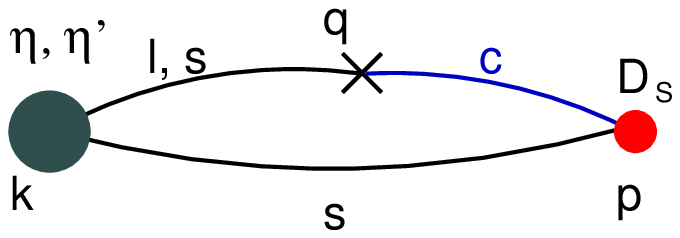}
 \hfil\hfil\includegraphics[width=0.3\linewidth]{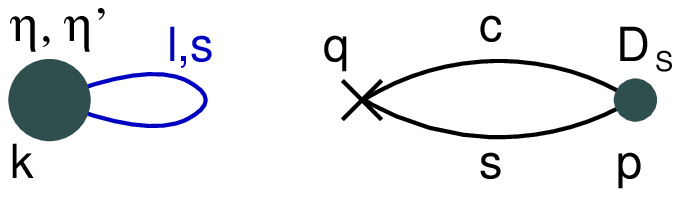}\\[-2.5em]
 \caption{Connected (left) and disconnected (right) diagrams which
 contribute to $C_3(t_f,t;\V{p},\V{q})$.  We use a stochastic method to
 estimate the all-to-all propagators, denoted by blue lines. }
 \label{fig:diagrams}
\end{figure}

In particular, the $D_s$ meson is interesting for flavor physics.
Its major semi-leptonic decay is to $\eta$ and $\eta'$, which has
a contribution from a disconnected loop diagram (Fig.~\ref{fig:diagrams}).  
The loop runs over three light
flavors so the effect is enhanced by a factor three, and thus may be large.
The purpose of this work is to test the feasibility of measuring
the disconnected diagram, and to quantify its contribution
to the form factor.

We extract the scalar form factor $f_0$ from the relation~\cite{Na:2009au}:
\begin{equation}
 f_0(q^2)
 =\frac{m_c - m_l}{m_{D_s}^2 - m_\eta^2}\langle \eta |S| D_s \rangle,
\label{eq:f0}
\end{equation}
where $S= \bar{l}c $ is a scalar current made from charm and light quarks.
$m_i$ are the masses of the quarks and mesons.
The matrix element can be extracted from 
the following ratio of 3-point over 2-point
functions:
\begin{equation}
 \langle \eta(\V{k},t_i) |S(\V{q},t)| D_s(\V{p},t_f) \rangle
  = Z_\eta Z_{D_s} \frac{C_3(t_f-t_i,t-t_i; \V{p},\V{q})}
    { C_2^{\eta}(t-t_i;\V{k}) C_2^{D_s}(t_f-t;\V{p})}
  = Z_\eta Z_{D_s} R(\V{k},\V{q},\V{p},t-t_i,t_f-t_i),
   \label{eq:matrixelement}
\end{equation}
and similarly for $\eta'$.
For large $t_f-t_i$ and $t-t_i$ this ratio should approach a constant.
$Z_{\eta}$ and $Z_{D_s}$ are the overlap factors between the meson state and the
interpolating operator,
which can be extracted from the two point functions
$C_2^{\eta}(t-t_i;\V{k})$ and $C_2^{D_s}(t_f-t;\V{p})$, respectively.
The two point functions for $\eta$ and $\eta'$ also have a disconnected part,
however, at $m_{\rm PS} \simeq 445\ {\rm MeV}$ 
we expect its contribution to the mass to be small and we neglect 
it in this first exploratory study.

We use QCDSF $24^3 \times 48$ $n_f=2+1$ configurations \cite{Bietenholz:2011qq}.
So far we only use the ${\rm SU}(3)$ symmetric set
($\kappa_l=\kappa_s=0.1209$) with lattice spacing $a\simeq0.08 {\ \rm fm}$.
This was generated using the tree-level Symanzik-improved gluonic
action and non-perturbatively improved Wilson fermions with stout links in the
derivative terms (SLiNC action).
We use the same relativistic quark action for the (quenched) charm quark
with $\kappa_{\rm charm}=0.11$. Note that since we use the flavor ${\rm SU}(3)$
symmetric configurations, the disconnected contributions in the 
$D_s \to \eta$ 3-point function cancel, when we identify $\eta=\eta_8$.
The Chroma software package~\cite{Edwards:2004sx} is used for some of
the analysis.

\section{Noise Reduction techniques}
In order to calculate the disconnected loop, all-to-all propagators 
are required.  These are estimated using stochastic methods,
which involve performing $N$ inversions of the light quark
Dirac operator for each configuration; $N$ should be large enough
to give sufficiently small stochastic errors relative to the gauge noise.
For some quantities the stochastic noise dominates the overall
uncertainty and it is important to use 
efficient noise reduction techniques.

We measure the disconnected ``loop''
\begin{align*}
 C_1(t;\V{p})
 &=\sum_{\V{x},\V{x}',\V{x}''}e^{i\V{p}\cdot\V{x}}
    \tr \left[\gamma_5 \phi(\V{x},\V{x}') { M^{-1}(\V{x}',t;\V{x}'',t)}
     {\phi(\V{x}'',\V{x})} \right],
\end{align*}
where $M$ is the Dirac operator for a light quark and
$\phi$ is a smearing function.  The stochastic estimation of the all-to-all
propagator $M^{-1}(\V{x}',t;\V{x}'',t)$ involves 
the following approximation:
\begin{equation}
 M^{-1}
 =\frac{1}{N}\sum_{i=1}^N |s_i\rangle\langle \eta_i|
   +\mathcal{O}\left(\frac{1}{\sqrt{N}} \right),
\end{equation}
where $|\eta_i\rangle$ is a random noise vector and 
$|s_i\rangle = M^{-1} |\eta_i\rangle$.
We use $\frac{1}{\sqrt{2}}( \mathbb{Z}_2 + i \mathbb{Z}_2$) 
complex random numbers for the noise vector.
For each $i$ we need to smear both $|\eta_i\rangle$ and $|s_i\rangle$ 
($|\eta_i\rangle$ must be
smeared \emph{after} solving for $|s_i\rangle$) 
so we need $2N$ applications of the smearing operator.
This significantly increases the computer time needed to calculate the
disconnected loop.
Time dilution (partitioning) \cite{Bernardson:1993yg} is implemented: 
the noise vector is only non-zero on one or two time slices.

We test the following three noise reduction techniques.
\begin{description}
 \item[Spin dilution/partitioning]
	    This uses projected noise vectors on a single spinor component
	    and sums over the projections afterwards \cite{Bernardson:1993yg}:
	    \begin{equation}
	     \frac{1}{N}\sum_{a=1}^4\sum_{i=1}^N 
	      |s_i^{(a)}\rangle \langle \eta_i^{(a)}|,
	    \end{equation}
	    where $|\eta_i^{(a)}\rangle = P^{(a)}|\eta_i\rangle$ 
	    is the projected noise vector.
	    It requires $4N$ inversions 
	    but for some quantities
	    the stochastic error is reduced by a factor greater than $2$. 
	    In addition, we can reduce
	    the cost of smearing because the spin projection $P^{(a)}$
	    commutes with the smearing of our choice.  
	    A naive scaling gives $8N$ smearing
	    operations, but we only need $5N$ applications: 
	    $4N$ for $|s_i^{(a)}\rangle$
	    and $N$ for $|\eta_i\rangle$.
 \item[Hopping Parameter Acceleration (HPA)]\cite{Thron:1997iy}
	    This is based on the following identity
	    \begin{equation}
	     (\kappa D)^n M^{-1} 
	      = M^{-1}  -\kappa D - (\kappa D)^2 - \cdots -  (\kappa D)^{n-1},
	      \label{eq:hpa}
	    \end{equation}
	    where $\kappa D$ is the hopping part of the Dirac operator.
	    Note that the derivative  operator satisfies 
	    $\tr[ \gamma_5 \kappa D] = 0$ due to the spinor
	    structure so that this term only contributes to the noise.  
	    This means that
	    $(\kappa D)^2 M^{-1}$ represents an improved estimate of 
	    $M^{-1}$ (we call it $n=2$ HPA). 
	    As long as the smearing is diagonal in spinor space, this is
	    also true for the smeared all-to-all propagator.
 \item[Truncated Solver Method (TSM)] 
	    For some quantities the ultra violet modes dominate.
	    In these cases, using a small number of CG iterations in the
	    solver for the solution vector $|s_i\rangle$ provides a good
	    approximation, for example, to the disconnected loop
	    \cite{Collins:2007mh,Bali:2009hu}.
	    To arrive at an unbiased estimate, a correction term needs
	    to  be added to the truncated part: 
	    \begin{equation}
	     M^{-1} = 
	      \frac{1}{N_1}\sum_{i=1}^{N_1} 
	      |s_{\rm trunc,}{}_i \rangle\langle \eta_i|
	      +  \frac{1}{N_2} \sum_{j=N_1+1}^{N_1+N_2} 
	      |s_{\rm bias,}{}_j \rangle
	      \langle\eta_j|.
	      \label{eq:tsm}
	    \end{equation}
	    The first term uses
	    the truncated solution $|s_{\rm trunc,}{}_i \rangle$,
	    which is cheap to calculate
	    and typically causes the main part of 
	    the stochastic error.
	    The second term contains 
	    $|s_{\rm bias,}{}_j \rangle
	    = |s_{\rm conv,}{}_j \rangle -|s_{\rm trunc,}{}_j \rangle$, 
	    where $|s_{\rm conv,}{}_j \rangle$ is a converged solution.
	    $|s_{\rm conv,}{}_j \rangle$ is expensive, 
	    and only accounts for a small part of the stochastic error
	    if $|s_{\rm bias,}{}_j\rangle $ does not contribute
	    significantly to the observable.
	    Therefore, by tuning parameters 
	    --- $n$: number of CG-iterations for the truncated part, 
	    $N_1$: number of stochastic noises for the truncated part,
	    $N_2$: number of stochastic noises for the bias part ---
	    we can reduce the total calculation cost.
	    We use a CG solver for the truncated solutions and a BiCGstab
	    solver for the converged solutions.

\end{description}

\section{Comparisons}

We investigate the noise reduction techniques using one configuration.
We use Wuppertal smearing \cite{Gusken:1989ad} for the quarks, 
with parameters which are
tuned to minimize the contributions from the excited states to the
effective mass.

\begin{figure}
\vspace*{-0.5em}
\hfil
 \includegraphics[width=0.45\linewidth]{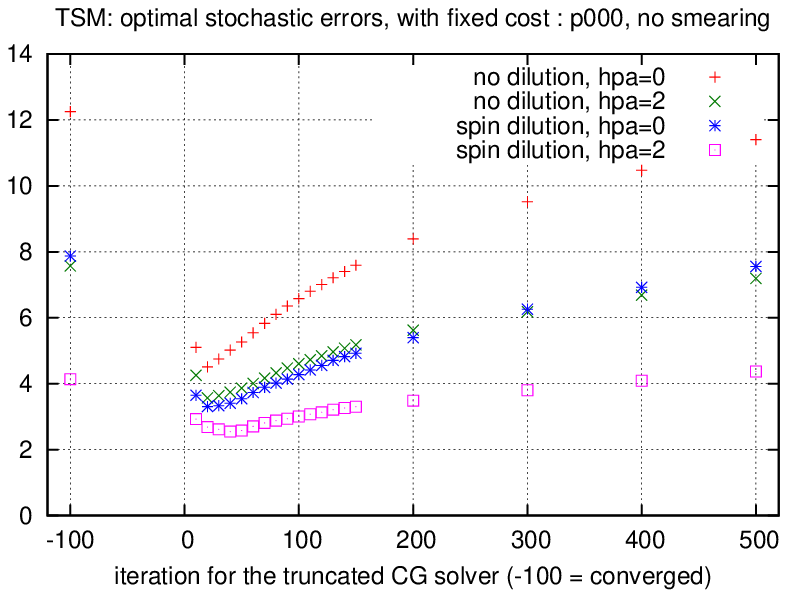}
\hfil
 \includegraphics[width=0.45\linewidth]{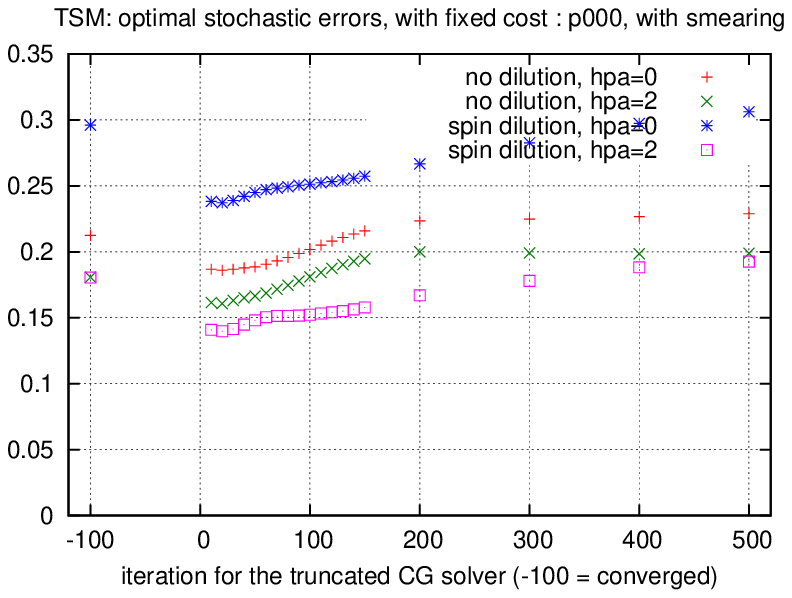}
\vspace{-0.5em}

\caption{Estimated stochastic errors at fixed cost for $\V{p}=(0,0,0)$.  
 The horizontal axes are $n$ for the TSM.  Data at $n=-100$ are without
 TSM.
 Left panel: without smearing.  
 Right panel: with smearing.}
 \label{fig:err-p000}
\end{figure}

\begin{figure}
\vspace{-0.5em}

\hfil
 \includegraphics[width=0.45\linewidth]{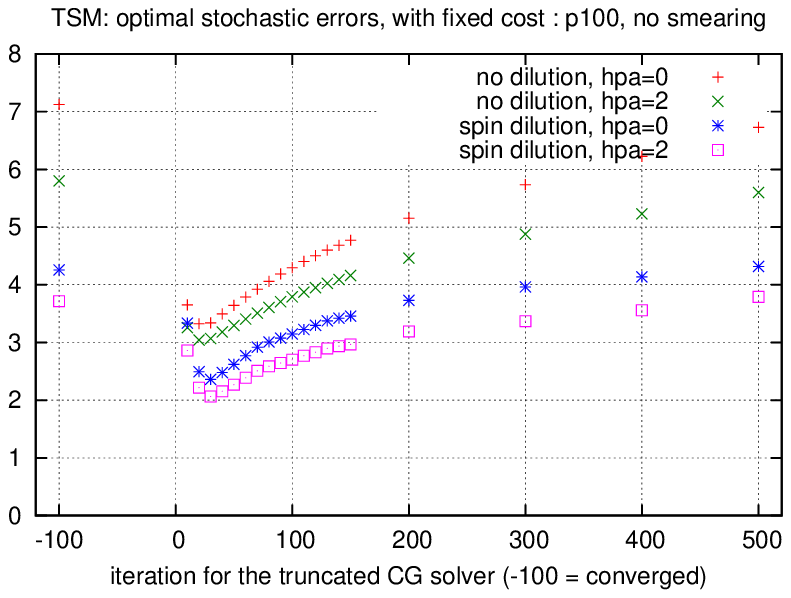}
\hfil
 \includegraphics[width=0.45\linewidth]{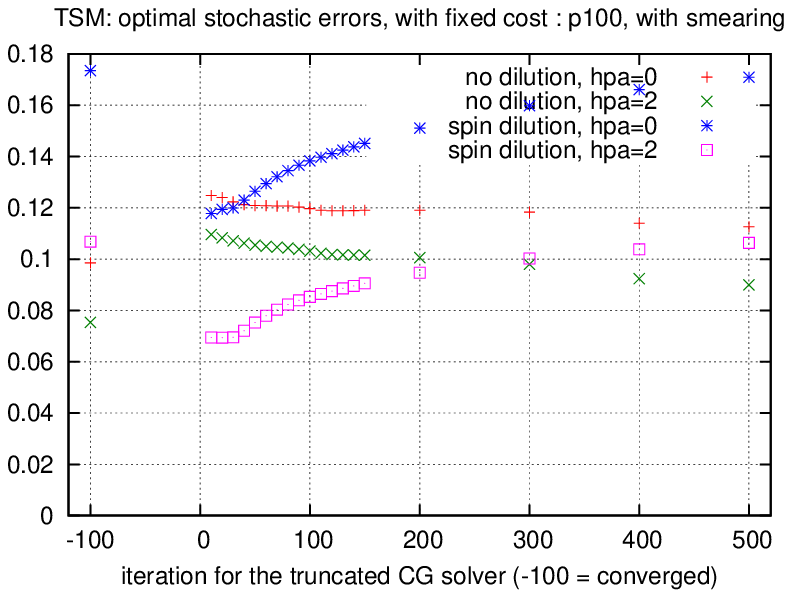}
\vspace{-0.5em}

\caption{The same as Fig.~\protect\ref{fig:err-p000} 
but for $\V{p}=(1,0,0)$.}
 \label{fig:err-p100}
\end{figure}

In Figs.~\ref{fig:err-p000} and \ref{fig:err-p100}
we plot the stochastic errors for
various combinations of the noise reduction techniques.
In each case, the computational cost is fixed.
The horizontal axes correspond to $n$, the number of iterations of
the solver in the TSM.
The data at $n=-100$ indicate the results without the TSM.
In particular, the red plus symbols (``+'') show the results 
without any noise reduction techniques.
For a fixed $n$, we have optimized $N_1$ and $N_2$ to give
the smallest stochastic error
under the cost condition
\begin{equation}
 N_1 ( n \tau_{\rm CG} + \tau_{\rm smear})
+ N_2( n \tau_{\rm CG} + n_{\rm conv} \tau_{\rm BiCGstab} + \tau_{\rm smear})
= \text{constant},
\end{equation}
assuming the square of error, $\sigma_{\rm stoch.}^2$, to scale according to
\begin{equation}
 \sigma^2_{\rm stoch.} = \frac{f_1}{N_1} + \frac{f_2}{N_2},
\end{equation}
where $f_1$ and $f_2$ are the variances of the first and second terms in
eq.~(\ref{eq:tsm}), respectively.
$n_{\rm conv}$ is the number of iterations
needed to obtain the converged solution.  $\tau_{\rm CG},\ 
\tau_{\rm BiCGStab}$ and $\tau_{\rm smear}$ represent
the computer time needed for 1 CG iteration, 1 BiCGstab iteration,
and smearing, respectively.
The optimal ratios of $N_1/N_2$ are around $1$ ($10$), 
with (without) smearing.

Although small differences between the results
are not significant due to the uncertainty on the stochastic errors,
in all cases spin dilution together with HPA (purple squares),
gives the minimum error when combined with TSM.
Therefore we use this combination in the following analysis.

The gain factor,
\begin{equation}
 g=
  \frac{\sigma^2(\text{without noise reduction})}
  { \sigma^2(\text{with noise reduction})},
\end{equation}
strongly depends on the smearing.
Without smearing (left panels), 
we obtain maximum gain factors of 16 -- 25,
which translates into a reduction of the computational cost 
of the same magnitude.
With smearing, it is only about a factor 2.
This is because the contribution to the error from the bias part (i.e.,
$f_2$) is larger than or of the same magnitude as the truncated part
($f_1$).


\section{Results}

Having optimized the noise reduction,
we can now measure the disconnected contribution to the 
form factor.  For the TSM, we truncate after $n=20$ CG iterations
and the numbers of noise vectors are $N_1=10$ and $N_2=20$.
A total of $939$ configurations were used in the analysis.

Following our previous study \cite{Evans:2010tg},
we use stochastic techniques for the connected contribution as well.
The noise vectors are placed at the sink of the 
$D_s$ meson (denoted by a red circle in Fig.~\ref{fig:diagrams}).
For each configuration,
$24\times 4$ spin diluted noise vectors were computed for the
charm quark.
In terms of momenta, $57$ different combinations of $\V{p}$ for the
$D_s$ meson were calculated.
Note that a similar calculation with the sequential method would require
$57\times 12$ inversions.

In order to extract the matrix elements in eq.~(\ref{eq:matrixelement}),
we fixed the time separation between the $\eta$ source
and the $D_s$ sink separately for the connected ($t_f=24$, $t_i=0$) 
and the disconnected ($t_f=24$, $t_i=16$) matrix elements.
We combine the two contributions afterwards.
For the connected part, taking the maximum separation $t_f-t_i=T/2=24$ 
enables us to average over the forward and backward propagations.
For the disconnected part, 
in order to average the forward and backward propagations, 
the noise vector has a non-zero value at two time-slices separated by 
$16$ time-slices ($t_f \pm 8$).
The usage of different $t_f-t_i$ for the connected and disconnected 3-point
functions is allowed because we have assumed
$m_\eta=m_{\eta'}$ (remember that $m_u=m_d=m_s$).

\begin{figure}
\vspace{-0.5em}

 \hfil
 \includegraphics[width=0.48\linewidth]{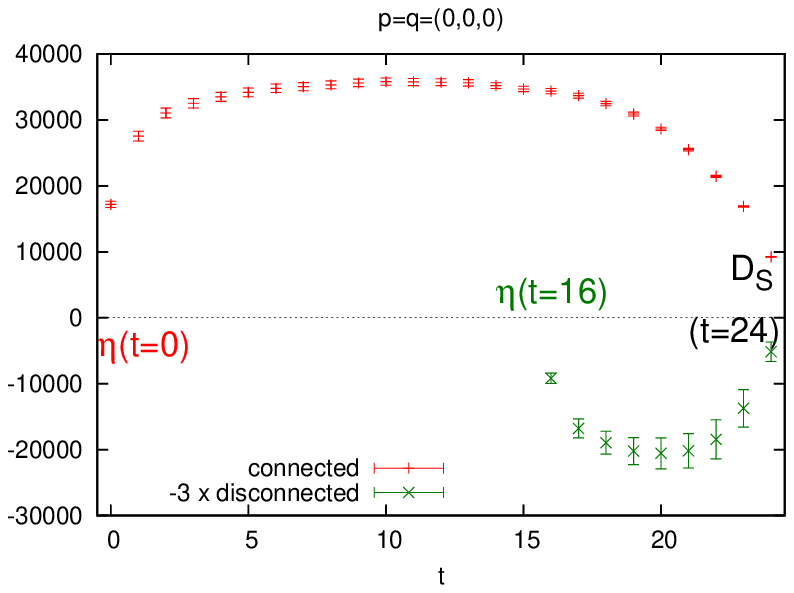}
 \hfil
 \includegraphics[width=0.48\linewidth]{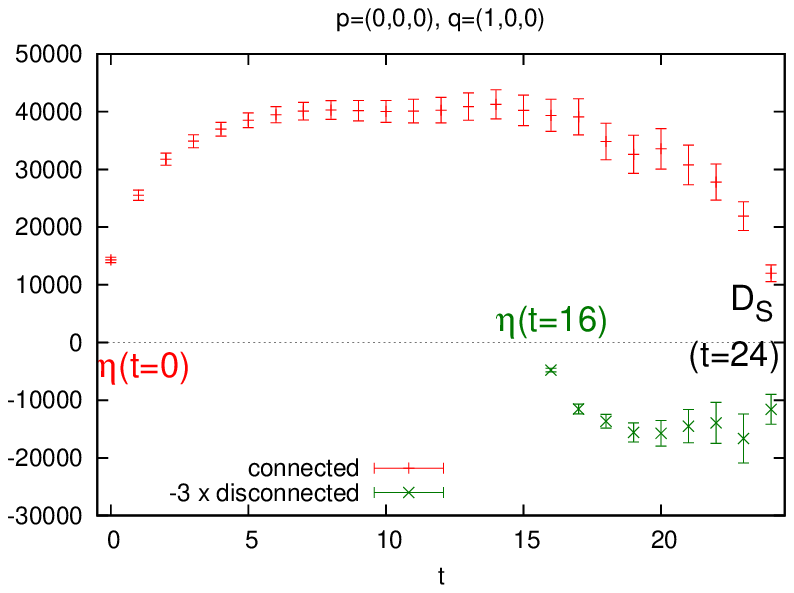}
 \caption{Ratios of 3-point over 2-point functions,
 $R$ in eq.~\protect\ref{eq:matrixelement},
for connected and
 disconnected parts.  
$\V{k}=\V{q}=(0,0,0)$ 
 for the left panel and $-\V{k}=\V{q}=(1,0,0)$ for the right panel.}
 \label{fig:plateau}
\end{figure}

Fig.~\ref{fig:plateau} shows 
the ratio of the correlation functions,
which corresponds to ${f_0(q^2)}/{Z_\eta Z_{D_s}}$.
The disconnected
part is multiplied by $3$ because of the $3$ light flavors.
The errors for the disconnected contribution
are small enough to obtain signals, significantly different from zero.

\begin{figure}
\vspace*{-0.5em}

 \hfil
 \includegraphics[width=0.56\linewidth
]{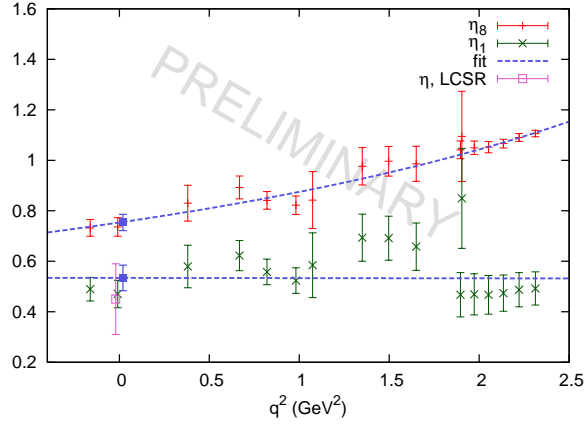}
\vspace*{-0.8em}

 \caption{
Form factor $f_0(q^2)$ for $D_s\to l\nu\eta_8$ and $D_s\to
 l\nu\eta_1$.  Errors are statistical only.
 A value from QCD light cone sum rules (LCSR) \cite{Azizi:2010zj} is also
 plotted.
 To enhance visibility, the fitted values at $q^2=0$ and the LCSR results
 are slightly shifted to the right and left, respectively.
}
\label{fig:formfactor}
\end{figure}

In Fig.~\ref{fig:formfactor} we show the form factors for
the octet ($\eta_8$) and singlet ($\eta_1$) $\eta$s:
\begin{align}
 |\eta_8 \rangle 
 &=\frac{1}{\sqrt{6}} 
    (|\bar{u}u\rangle + |\bar{d}d\rangle - 2|\bar{s}s\rangle)  
 &&\text{connected only,} \\
|\eta_1 \rangle 
 &=\frac{1}{\sqrt{3}} 
   (|\bar{u}u\rangle + |\bar{d}d\rangle + |\bar{s}s\rangle)
 && \text{connected${}-3\times\;$disconnected.}
\end{align}
Preliminary fits to $f_0(q^2)$ of the form $f_0(q^2)=\frac{f_0(0)}{1-b q^2}$ 
give $f_0(0) = 0.75(3)$ and $f_0(0) = 0.52(5)$, for $D_s \to \eta_8$
and $D_s \to \eta_1$,  respectively.
Also included in Fig.~\ref{fig:formfactor} is a value from light cone
QCD sum rules 
for the decay into $\eta$ \cite{Azizi:2010zj}, $f_0(0)=0.45(14)$.
Due to  ${\rm SU}(3)$ flavor symmetry 
the $D_s\to \eta_8$ form factor also represents
the form factor of $D\to l\nu \pi$ and $D\to l\nu K$.
Note that $f_0(0)$ for $\eta_1$ is smaller than that for $\eta_8$.
This is consistent with the form factors for 
$B\to \eta,\eta'$ \cite{Ball:2007hb},
which is the heavy quark limit, can be compared to our calculation.


\section{Conclusions}

We tested three methods (and their combinations) of noise reduction
techniques for measuring the disconnected contributions to the $D_s$ meson 
semi-leptonic decay form factor.
The combination of spin dilution,
hopping parameter acceleration and truncated solver method was found to
give the biggest gain in computer time.
These noise reduction techniques
allowed us to measure non-zero contributions to the form factor, 
on ${\rm SU}(3)$ flavor symmetric QCDSF $n_f=2+1$ configurations.
Further studies with non-${\rm SU}(3)$ symmetric $n_f=2+1$
configurations are planned.

\medskip

This work was supported by the EU ITN STRONGnet (grant number 238353)
and the DFG SFB/Transregio 55.  SC 
acknowledges support from the Claussen-Simon-Foundation (Stifterverband
f\"ur die Deutsche Wissenschaft).
JZ is supported by
the Australian Research Council under grant FT100100005.
The calculations were performed on the Athene HPC cluster at
the University of Regensburg.

\end{document}